\documentclass[pra,aps,showpacs,twocolumn,superscriptaddress, floatfix]{revtex4}
\usepackage{graphicx}
\usepackage[ansinew]{inputenc}
\usepackage{array}
\usepackage{color}
\usepackage{amsmath}
\usepackage{amsxtra}
\usepackage{amstext}
\usepackage{amssymb}
\usepackage{latexsym}
\begin{document}
\title{Dissipation-driven two-mode mechanical squeezed states in optomechanical systems}
\author{Huatang Tan}
\affiliation{Department of physics, Huazhong Normal University, Wuhan 430079, China}
\affiliation{B2 Institute, Department of Physics and College of Optical Sciences,
The University of Arizona, Tucson, AZ 85721}
\author{Gaoxiang Li}
\affiliation{Department of physics, Huazhong Normal University, Wuhan 430079, China}
\author{P. Meystre}
\affiliation{B2 Institute, Department of Physics and College of Optical Sciences,
The University of Arizona, Tucson, AZ 85721}
\begin{abstract}
In this paper, we propose two quantum optomechanical arrangements that permit the dissipation-enabled generation of steady two-mode mechanical squeezed states. In the first setup, the mechanical oscillators are placed in a two-mode optical resonator while in the second setup the mechanical oscillators are located in two coupled single-mode cavities. We show analytically that for an appropriate choice of the pump parameters, the two mechanical oscillators can be driven by cavity dissipation into a stationary two-mode squeezed vacuum, provided that  mechanical damping is negligible. The effect of thermal fluctuations is also investigated in detail and show that ground state pre-cooling of the oscillators in not necessary for the two-mode squeezing. These proposals can be realized in a number of optomechanical systems with current state-of-the-art experimental techniques.
\end{abstract}

\maketitle
\section{Introduction}
Quantum squeezing and entanglement have been observed in a number of atomic and photonic systems and are expected to play an increasing role in applications ranging from measurements of feeble forces and fields to quantum information science~\cite{qm1, qm2, qi1, qi2}.  For example, it has been know for over three decades that squeezed vibrational states are of importance for the measurement beyond the standard quantum limit of the weak signals expected to be produced e.g. in  gravitational wave antennas~\cite{fm}.

Although achieving such effects in macroscopic systems remains a major challenge due in particular to the increasing rate of environment-induced decoherence~\cite{dec}, recent progress toward the ground-state cooling of micromechanical systems \cite{c1, c2, c3, c4,c5} may change the situation significantly in the near future. In particular, the characterization of quantum ground-state mechanical motion \cite{qzm1}, the quantum control of a mechanical resonator deep in the quantum regime and coupled to a quantum qubit \cite{qzm2}, and the demonstration of optomechanical ponderomotive squeezing~\cite{pm} are important steps toward the broad exploration of quantum effects in truly macroscopic systems~\cite{mq1, mq2, mq3, mq4, mq5, mq6, mq7, mq9,opm}.  Several proposals have been put forward to generate mechanical squeezing in an optomechanical oscillator, including the injection of nonclassical light~\cite{ms1}, conditional quantum measurements~\cite{ms2}, and parametric amplification \cite{ms3, ms4, ms5, ms6}. In all cases, decoherence and losses are a dominant limiting factor in the amount of squeezing that can be achieved.

A new paradigm in quantum state preparation and control has recently received increased attention. Its key aspect is that it exploits quantum dissipation in the generation of specific quantum states. Quantum reservoir engineering has been proposed to prepare desirable quantum states \cite{re1, re2, re3, re4} and perform quantum operations \cite{re5}, and the creation of steady-state entanglement between two atomic ensembles by quantum reservoir engineering has been experimentally demonstrated \cite{re6}. This dissipative approach to quantum state preparation presents the double advantage of being independent of specific initial states and of leading to steady states robust to decoherence.

Here, we propose to exploit cavity dissipation to generate the steady-state two-mode squeezing of two spatially separated mechanical oscillators. These states are also entangled states of continuous variables, a basic resource in quantum information precessing. We propose specifically two different setups. In the first one two mechanical oscillators are placed inside a two-mode optical resonator, while in the second one the mechanical oscillators are located in two separate single-mode cavities coupled by photon tunneling. The cavities are driven in both cases by amplitude-modulated lasers. In the first  model we show analytically that for appropriate mechanical oscillator positions and pump laser parameters the mechanical oscillators can be driven into a stationary two-mode squeezed vacuum by cavity dissipation, provided that mechanical damping is negligible. In the second case a two-step driving sequence can likewise give rise to a steady two-mode mechanical squeezed vacuum state. The effect of thermal fluctuations on the resulting  squeezed states is also investigated in detail.

The remainder of this paper is organized as follows. Section II introduces our first model of two mechanical oscillators coupled by radiation pressure to the field of a common two-mode optical resonator. It then discusses how to prepare a two-mode mechanical squeezed state. Section III shows that cavity damping can likewise be exploited to create a squeezed mechanical state of two mechanical oscillators in separated single-mode cavities. Finally, Section IV is a conclusion and outlook.

\section{Mechanical oscillators in a single two-mode cavity}
\subsection{Model and equations}
We first consider an extension of the ``membrane-in-the-middle" arrangement of cavity optomechanics~\cite{mid1} where two mechanical oscillators, modeled as vibrating dielectric membranes of identical frequencies $\omega_m$, are located inside a driven two-mode optical resonator, as illustrated in Fig.~1. The mechanical modes are characterized by the bosonic annihilation operators $\hat C_j$ and the cavity modes, at frequencies $\omega_{cj}~(j=1,2)$, by annihilation operators $\hat A_j$. These modes are driven by amplitude-modulated lasers of frequencies $\omega_{lj}$. The Hamiltonian of the driven cavity-oscillators system reads ($\hbar=1$)
\begin{eqnarray}
H&=&\sum_{j,k=1,2}\Big\{\omega_{cj} \hat A_j^\dag \hat A_j+\omega_m \hat C_j^\dag \hat C_j+g_{jk}\hat A_j^\dag \hat A_j(\hat C_k+\hat C_k^\dag)\nonumber\\
&+&i\mathcal{E}_j(t)e^{-i\omega_{lj}t} \hat A_j^\dag-i\mathcal{E}_j^*(t)e^{i\omega_{lj}t} \hat A_j\Big\}, \label{hamil}
\end{eqnarray}
where $\mathcal{E}_j(t)$ are the time-dependent amplitudes of the pump lasers. Their specific forms will be given later. The single-photon optomechanical coupling constants $g_{jk}=\frac{\omega_{cj} f_{jk}(\bar{x}_k)}{L\sqrt{m\omega_m}}$
\cite{mid1}, where $L$ is the length of the cavity, $m$ are the identical masses of the membranes, and
$f_{jk}(\bar{x}_k)=\frac{2\mathcal{R}_k\sin(2k_{cj}\bar{x}_{k})}
{\sqrt{1-\mathcal{R}_k^2\cos^2(2k_{cj}\bar{x}_k)}}$.
Here $\mathcal{R}_k$ and $\bar{x}_{k}$ are the reflection coefficients and equilibrium positions of the two membranes and $k_{cj}$ the wave numbers of the cavity modes.

For an appropriate combination of cavity length and membrane positions of the membranes, see Fig.1, it is possible to find a situation such that  the ``symmetrical'' and ``antisymmetric'' optomechanical coupling strengths satisfy  the equalities
\begin{eqnarray}
g_{11}&=&g_{12}=g_{1},\nonumber \\
g_{21}&=&-g_{22}=g_{2},
\end{eqnarray}
as discussed in Ref.~\cite{mid2}. Introducing then the new bosonic annihilation  operators
\begin{subequations}
\label{bm}
\begin{align}
\hat B_1&=(\hat C_1+\hat C_2)/{\sqrt{2}},\\
\hat B_2&=(\hat C_1-\hat C_2)/{\sqrt{2}},
\end{align}
\end{subequations}
the Hamiltonian $H$ becomes $H=\sum_j\tilde{H}_j$, where
\begin{eqnarray}
\tilde{H}_j&=&\omega_{cj} \hat A_j^\dag \hat A_j+\omega_{m}B_j^\dag \hat B_j+g_j\hat A_j^\dag \hat A_j(\hat B_j+\hat B_j^\dag)\nonumber\\
&+&i\mathcal{E}_j(t)e^{-i\omega_{lj}t} \hat A_j^\dag-i\mathcal{E}_j^*(t)e^{i\omega_{lj}t} \hat A_j.
\end{eqnarray}
In terms of the normal operators $\hat B_j$, the system is therefore decoupled and  reduces to two independent single-membrane optomechanical systems.
\begin{figure}[t]
\includegraphics[width=3.0in]{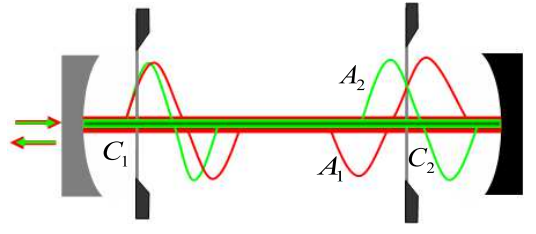}
\caption{Schematic plot of two vibrating membranes ($C_j$) placed at appropriately-chosen positions in a driven cavity with two frequency-nondegenerate resonant modes ($A_j$).}
\end{figure}

Further decomposing the operators $\hat A_j$ and $\hat B_j$ as
\begin{eqnarray}
\hat A_j&=& \langle \hat A_j(t)\rangle+\hat a_j \equiv \alpha_j(t)+\hat a_j,\nonumber \\
 \hat B_j&=&\langle \hat B_j(t)\rangle + \hat b_j \equiv \beta_j(t)+\hat b_j,
\end{eqnarray}
and with  $|\alpha_j(t)|^2\gg\langle \hat{a}_j^\dag \hat{a}_j\rangle$ and $|\beta_j(t)|^2\gg\langle \hat{b}_j^\dag \hat{b}_j\rangle$ one can linearize the Hamiltonians $\tilde{H}_j$ to get
\begin{equation}
\tilde{H}_j^{\rm lin}=\Delta_j \hat a_j^\dag \hat a_j+\omega_m \hat b_j^\dag \hat b_j+[\chi_j^*(t)\hat a_j+\chi_j(t)\hat a_j^\dag] (\hat b_j+\hat b_j^\dag),\label{hamil1}
\end{equation}
where
$$
\Delta_j=\delta_j+g_j(\beta_j+\beta_j^*),
$$
with $\delta_j=\omega_{cj}-\omega_{lj}$, and the effective optomechanical coupling strengths are given by
\begin{equation}
\chi_{j}(t)=g_j\alpha_j(t).
\label{effective g}
\end{equation}

The density matrix $\tilde{\rho}_j$ of the sub-system composed of the cavity mode $a_j$ and the normal mode $b_j$ satisfies the master equation
\begin{eqnarray}
\label{master1}
\dot{\tilde{\rho}}_j(t)&=&-i[\tilde{H}_j^{\rm lin}, \tilde{\rho}_j]+\frac{\kappa_j}{2}(2\hat a_j\tilde{\rho}_j \hat a_j^\dag- \hat a_j^\dag \hat a_j\tilde{\rho}_j- \tilde{\rho}_j \hat a_j^\dag \hat a_j)\nonumber\\
&+&\frac{\gamma_m}{2}(\bar{n}_{\text{th}}+1)(2\hat b_j\tilde{\rho}_j \hat b_j^\dag- \hat b_j^\dag \hat b_j\tilde{\rho}_j- \tilde{\rho}_j \hat b_j^\dag \hat b_j)\nonumber\\
&+&\frac{\gamma_m}{2}\bar{n}_{\text{th}}(2\hat b_j^\dag\tilde{\rho}_j \hat b_j- \hat b_j \hat b_j^\dag\tilde{\rho}_j- \tilde{\rho}_j \hat b_j \hat b_j^\dag ),
\end{eqnarray}
where $\kappa_j$ is the cavity dissipation rate and $\gamma_m$ the mechanical damping rate, taken to be the same for both oscillators. The mean thermal phonon number is $\bar{n}_{\text{th}}=(e^{\hbar\omega_m/k_BT}-1)^{-1}$, with $k_B$ the Boltzmann constant and $T$ the temperature.

\subsection{Stationary two-mode mechanical squeezed vacuum via cavity dissipation\label{ii}}

We now show how cavity dissipation can be exploited to prepare the stationary two-mode mechanical squeezed vacuum of the vibrating membranes. To this end we consider an effective optomechanical coupling strength $\chi_j(t)$ of the form
\begin{equation}
\label{cp1}
\chi_j(t)=\chi_{j1}e^{-i(\Omega_j t-\phi_j)}+\chi_{j2},
\end{equation}
where $\chi_{j1}$ and $\chi_{j2}$ are constants. This situation can be realized by a pump laser of the form discussed in Section~\ref{iic}.

For  weak optomechanical coupling we have $\Delta_j\simeq \delta_j$, so that if the cavity-laser detuning $\delta_j$ and the modulation frequency $\Omega_j$ are
\begin{equation}
\delta_j=\omega_m,\,\,\,\,\,\,\,\,\,\,\Omega_j=2\omega_m,
\end{equation}
the transformation $\hat a_j\rightarrow \hat a_je^{-i\Delta_jt}$ and $\hat b_j\rightarrow \hat b_je^{-i\omega_mt}$ reduces the Hamiltonian $\tilde{H}_j^{\rm lin}$  to
\begin{eqnarray}
\tilde{H}_j^{\rm lin}&=&(\chi_{j1} e^{-i\phi_j} \hat b_j+\chi_{j2} \hat b_j^\dag)a_j\\
&+&(\chi_{j1} e^{2i\omega_m t-i\phi_j} \hat b_j+\chi_{j2}e^{-2i\omega_m t} \hat b_j^\dag) \hat a_j+\text {H.c.} \nonumber
\end{eqnarray}
This form can be further simplified for a mechanical frequency $\omega_m\gg \chi_{jk}$, in which case the rapid oscillating terms $e^{\pm2i\omega_mt}$ can be neglected and
\begin{equation}
\label{ham2}
\tilde{H}_j^{\rm lin}\simeq(\chi_{j1} e^{-i\phi_j} \hat b_j+\chi_{j2} \hat b_j^\dag) \hat a_j+\text {H.c.}
\end{equation}
This Hamiltonian describes the coupling of the cavity field $a_j$ to the normal mode $b_j$ simultaneously via parametric amplification -- through the term proportional to $\chi_{j1}$-- and via a beam splitter-type coupling -- through the constant contribution $\chi_{j2}$. It is known that the first term in the Hamiltonian~({\ref{ham2}) leads to photon-phonon entanglement and phononic heating of the normal mode $b_j$, while the second term results in quantum state transfer between the cavity mode $a_j$ and the mechanical mode $b_j$ as well as to mechanical cooling (cold damping). To ensure the stability of the system, the coupling strengths must satisfy the inequality $\chi_{j2}>\chi_{j1}$, that is, cooling should dominate over anti-damping.

When absorbing a laser photon the parametric amplification term results in the simultaneous emission of a photon into the cavity mode $a_j$ and a phonon in the normal mode $b_j$, while the beam-splitter interaction corresponds to the annihilation of a photon and the emission of  a phonon. We now show that the combined effect of these processes is the generation of steady-state single-mode squeezing of the normal mechanical mode $b_j$, or equivalently of two-mode squeezing of the original modes $c_1$ and $c_2$.

We proceed by first performing the unitary transformation
\begin{equation}
\label{ut}
{\varrho}_j(t)=\hat S_j^\dag(\xi_j)\tilde{\rho}_j(t)\hat S_j(\xi_j),
\end{equation}
where the squeezing operator
\begin{equation}
\hat S(\xi_j)=\exp \left [ -{\xi_j^* \hat b_j^{\dag2}/2+\xi_j}\hat b_j^2/2\right ],
\end{equation}
with
\begin{eqnarray}
\xi_j&=&r_j e^{-i\phi_j}, \\
r_j&=&\tanh^{-1}(\chi_{j1}/\chi_{j2}).
\end{eqnarray}
For $\gamma_m=0$ the master equation~(\ref{master1}) becomes then
\begin{equation}
\dot{\varrho}_j(t)=-i[{\cal H}_j^{\rm lin}, \varrho_j]+\frac{\kappa_j}{2}(2\hat a_j\varrho_j \hat a_j^\dag- \hat a_j^\dag \hat a_j\varrho_j- \varrho_j \hat a_j^\dag \hat a_j), \label{master3}
\end{equation}
where
\begin{equation}
{\cal H}_j^{\rm lin}=G_j( \hat a_j^\dag \hat b_j+ \hat b_j^\dag  \hat a_j),
\end{equation}
with
\begin{equation}
G_j=\chi_{j2}\sqrt{1-(\chi_{j1}/\chi_{j2})^2},
\label{G}
\end{equation}
describes quantum-state transfer between the cavity mode $a_j$ and the mechanical mode $b_j$ in the transformed picture.
It can be inferred from that master equation that in the transformed picture and for $\gamma_m=0$ the cavity mode $a_j$ and the mechanical mode $b_j$ asymptotically decay to the ground state
\begin{equation}
\varrho_j(\infty)=|0_{a_j}0_{b_j}\rangle\langle 0_{a_j}0_{b_j}|.
\end{equation}
Simply reversing the unitary transformation~(\ref{ut}) we have then that in the steady-state regime the normal mode $b_j$ is indeed in the squeezed vacuum state
\begin{equation}
\label{squ1}
\tilde{\rho}_{b_j}(\infty)=\hat S(\xi_j)|0_{b_j}\rangle\langle0_{b_j}|\hat S^\dag(\xi_j).
\end{equation}

It is possible to adjust the amplitude $\chi_{jk}$ and phase $\phi_j$ of the optomechanical coupling coefficient ~(\ref{cp1}) in such a way that the squeezing parameters satisfy the conditions
\begin{subequations}
\begin{gather}
r \equiv r_j,\\
\phi \equiv \phi_1=\phi_2-\pi, \label{phase}
\end{gather}
\label{specific}
\end{subequations}
in which case the two normal modes $b_1$ and $b_2$ exhibit the same amount of steady-state squeezing, but in perpendicular directions. With Eqs.~(\ref{bm}) and (\ref{squ1}) we then have
\begin{equation}
\rho_{c_1c_2}(\infty)=\hat S_{12}(\xi_{12})|0_{c_1},0_{c_2}\rangle\langle0_{c_1},0_{c_2}| \hat S_{12}^\dag(\xi_{12})
\label{2ms}
\end{equation}
where we have introduced the two-mode squeezing operator
\begin{equation}
\hat S_{12}(\xi_{12})=\exp (-\xi_{12}^*\hat c_1^\dag \hat c_2^\dag+\xi_{12}\hat c_1\hat c_2),
\end{equation}
and $\xi_{12}=r e^{-i\phi}$.
This two-mode squeezed vacuum of the mechanical oscillators can be thought of as the output from a 50:50 beam splitter characterized by the unitary transformation~(\ref{bm}), the two inputs being the normal modes $b_j$ squeezed in perpendicular directions. This shows that for $\gamma_m \rightarrow 0$ the dissipation of the intracavity field can be exploited to prepare pure two-mode mechanical squeezed vacuum state. The minimum time $t_{\rm min}$ required for preparing such states can be evaluated from the eigenvalues of Eq~(\ref{master3}),
$$
\eta_{j\pm}=-\frac{\kappa_j}{2}\pm\sqrt{\frac{\kappa_j^2}{4}-G_j^2}.
$$ For symmetric parameters $\kappa=\kappa_j$, $\chi_j=\chi_{jk}$, and $G_j=G$ one finds $t_{\rm min}=4/\kappa$ for $G\ge \kappa/2$.

\subsection{The choice of pump lasers \label{iic}}
Next we consider the choice of pump laser amplitudes $\mathcal{E}_j(t)$ that result in the time-dependent optomechanical coupling~(\ref{cp1}), and the specific form~(\ref{specific}) that results in the pure two-mode mechanical squeezing~(\ref{2ms}). From Eq.~(\ref{effective g}) an obvious starting ansatz is
\begin{equation}
\mathcal{E}_j(t)={\cal E}_{j1}e^{-i(\Omega_jt -\varphi_{j1})}+{\cal E}_{j2}e^{i\varphi_{j2}},
\end{equation}
where $\varphi_{jk}$ are the initial phases of two components with amplitudes ${\cal E}_{jk}$ and $\Omega_j$ are modulation frequencies. The corresponding amplitudes $\alpha_j(t)$ and $\beta_j(t)$ of the cavity and mechanical modes are
\begin{eqnarray}
\frac{d}{dt}\alpha_j(t)&=&-[\kappa+i\delta_j+ig_j(\beta_j+\beta_j^*)]\alpha_j +{\cal E}_{j1}e^{-i(\Omega_jt-\varphi_{1j})}\nonumber\\
&+&{\cal E}_{j2}e^{i\varphi_{j2}},\nonumber \\
\frac{d}{dt}\beta_j(t)&=&-(\gamma_m+i\omega_m)\beta_j-ig_j|\alpha_j|^2.
\label{cme}
\end{eqnarray}
It is difficult to find exact solutions of these equations in general. For the case of the weak optomechanical coupling strengths $g_j$, however, approximate analytical solutions can be found by expanding the amplitudes $\alpha_j$ and $\beta_j$ in powers of $g_j$ as
$\alpha_j=\alpha_j^{(0)}+\alpha_j^{(1)}+\alpha_j^{(2)}+\cdots$ and $\beta_j=\beta_j^{(0)}+\beta_j^{(1)}+\beta_j^{(2)}+\cdots$. Substituting these into Eqs.~(\ref{cme}) and for the time $t\gg 1/\kappa$, $\omega_m\gg\gamma_m$, $\delta_j=\omega_m$, and $\Omega_j=2\omega_m$ one then finds (higher-order corrections can be derived straightforwardly)
\begin{subequations}
\begin{align}
\alpha^{(0)}_j&=\frac{\varepsilon_{j1}}{\sqrt{\kappa^2+\omega_m^2}}e^{-i(\Omega_jt-\phi_j)}
+\frac{\varepsilon_{j2}}{{\sqrt{\kappa^2+\omega_m^2}}},\\
\alpha^{(1)}_j&=0,~~\beta^{(0)}_j=0,~~\beta^{(2)}_j=0,\\
\alpha^{(2)}_j&=\frac{2ig_j^2 {\cal E}_{j2}(2 {\cal E}_{j1}^2+3 {\cal E}_{j2}^2)
}{3\omega_m(\kappa^2+\omega_m^2)^2}e^{-i(\phi_j-\varphi_{j1})}\nonumber\\
&~+\frac{2ig_j^2{\cal E}_{j1}(3 {\cal E}_{j1}^2+2 {\cal E}_{j2}^2)}{3\omega_m(\kappa^2+\omega_m^2)^2}
e^{-i\left [\Omega_jt- (2\phi_j-\varphi_{j1})\right ]}\nonumber\\
&~-\frac{2ig_j^2 {\cal E}_{j1} {\cal E}_{j2}^2}{3\omega_m(\kappa^2+\omega_m^2)^2}
e^{i(\Omega_jt-\varphi_{j1})}\nonumber\\
&~-\frac{2ig_j^2 {\cal E}_{j1}^2 {\cal E}_{j2}}{3\omega_m(\kappa^2+\omega_m^2)^{3/2}(\kappa-3i\omega_m)}e^{-2i(\Omega_jt - \phi_j)},\\
\beta^{(1)}_j&=-\frac{g_j({\cal E}_{j1}^2+{\cal E}_{j2}^2)}{\omega_m(\kappa^2+\omega_m^2)}
-\frac{g_j{\cal E}_{j1}{\cal E}_{j2}}{3\omega_m(\kappa^2+\omega_m^2)}
e^{i(\Omega_jt-\phi_j)}\nonumber\\
&~+\frac{g_j{\cal E}_{j1}{\cal E}_{j2}}{\omega_m(\kappa^2+\omega_m^2)}
e^{-i(\Omega_jt-\phi_j)},
\end{align}
\end{subequations}
with the phases
\begin{eqnarray}
\phi_j&=&\varphi_{j1}+ \varphi_{j2},\nonumber \\
\varphi_{j2}&=&\arctan(\omega_m/\kappa).
\label{phase1}
\end{eqnarray}

For coupling strengths $g_j\ll\{\kappa, \varepsilon_{jk},\omega_m\}$, for example for $g_j\sim10^{-6}\omega_m$, $\kappa\sim0.05\omega_m$, and ${\cal E}_{jk}\sim 10^4\omega_m$, we find $|\alpha_j^{(2)}|\sim1\ll |\alpha_j^{(0)}|\sim10^4$ and $\delta_j\sim\omega_m\gg g_j(\beta_j+\beta_j^*)\sim 10^{-4}\omega_m$. We can therefore set $\Delta_j=\delta_j+g_j(\beta_j+\beta_j^*)\simeq\delta_j$, and the effective optomechanical coupling strength $\chi_j(t)\simeq g_j\alpha_j^{(0)}$ takes the required form~(\ref{cp1}) for
\begin{equation}
\chi_{jk}=\frac{g_j{\cal E}_{jk}}{\sqrt{\kappa^2+\omega_m^2}}.
\end{equation}
Finally, Eqs.~(\ref{phase1}) and~(\ref{phase}) give
\begin{equation}
\varphi_{21}-\varphi_{11}=\pi,~~\varphi_{j2}=\arctan(\omega_m/\kappa).
\end{equation}

\subsection{Thermal fluctuations}

For finite mechanical damping, $\gamma_m \neq 0$, the mechanical modes $c_1$ and $c_2$ are no longer be in a pure squeezed state, but rather in a two-mode squeezed thermal state. In the state, we find from Eqs.~ (\ref{bm}) and (\ref{master1})
\begin{subequations}
\label{meanv}
\begin{align}
\langle \hat c_j^\dag \hat c_j\rangle_\infty&=d_0 d_1\cosh2r-d_0d_2\sinh2r+\sinh^2r,\\
\langle \hat c_1 \hat c_2\rangle_\infty&=\left [-(d_0d_1+1/2)\sinh2r+d_0d_2\cosh2r \right] e^{i\phi},
\end{align}
\end{subequations}
where $d_1=\bar{n}_{\text{th}}\cosh2r+\sinh^2r$, $d_2=(\bar{n}_{\text{th}}+\frac{1}{2})\sinh2r$, and
\begin{equation}
d_0=1-\frac{4\kappa G^2}{(\kappa+\gamma_m)(\kappa\gamma_m+4G^2)}.
\label{para-d}
\end{equation}

The quantum correlations between the mechanical oscillators can be quantified by the sum of variances \cite{qi1, duan}
\begin{equation}
\Delta_{\rm EPR}=\langle(\hat X_1^{\theta_1}+\hat X_2^{\theta_2})^2\rangle +\langle(\hat X_1^{\theta_1+\frac{\pi}{2}}-\hat X_2^{\theta_2+\frac{\pi}{2}})^2\rangle,
\label{delta epr}
\end{equation}
where $\hat X_j^{\theta_j}=( \hat c_j e^{-i\theta_j}+ \hat c_j^\dag e^{i\theta_j})/\sqrt{2}$ are quadrature operators with local phase $\theta_j$. A value of $\Delta_{\rm EPR}<2$ is a signature of Einstein-Podolsky-Rosen (EPR)-type correlations between the two mechanical modes, with $\Delta_{\rm EPR}=0$ corresponding to the ideal quantum mechanical limit \cite{qi1}.  In our case we find that $\Delta_{\rm EPR, min}$ is minimum for the choice of local phases $\theta_1+\theta_2=\phi$. In the long-time limit it is equal to
\begin{equation}
\Delta_{\rm EPR, min}=2e^{-2r}(1-d_0)+2(2\bar{n}_{\text{th}}+1)d_0.
\label{epr min}
\end{equation}

We first observe that for $\kappa=0$ we have $d_0=1$ and hence $\Delta_{\rm EPR}(\infty)=4\bar{n}_{\text{th}}+2$, showing that in the absence of cavity dissipation the oscillators are just prepared in the thermal state imposed by their mechanical coupling to the environment. This confirms that cavity dissipation is an essential component of this realization of steady-state two-mode mechanical squeezing.
From Eq.~(\ref{epr min}), steady-state squeezing requires that
\begin{equation}
\bar{n}_{\text{th}}<\bar{n}_{\rm th, max}=\frac{1-d_0}{2d_0}\left (1-e^{-2r}\right ).\label{ineq1}
\end{equation}
In practice, it is important to be able to operate at as high a number of mean thermal phonons as possible. This can be achieved by decreasing $d_0$ while keeping the ``cooperative parameter'' $\frac{4G^2}{\kappa\gamma_m}\gg1$. From Eq.~(\ref{para-d}) and for the realistic case $\kappa \gg \gamma_m$,  $d_0$ reduces approximately to
\begin{equation}
d_0\simeq \frac{\gamma_m}{\kappa}+\frac{\kappa\gamma_m}{4G^2},
\end{equation}
 indicating that by increasing the coupling frequency $G$ while keeping the ratio $\chi_1/\chi_2$ fixed it is possible to increase the value of $\bar{n}_{\rm th, max}$ which is approximately given by
\begin{equation}
\bar{n}_{\rm th, max}\simeq\frac{4\kappa \chi_1(\chi_2-\chi_1)}{\gamma_m\left [\kappa^2+4(\chi_2^2-\chi_1^2)\right ]}.
\end{equation}
As a concrete example, for a cavity dissipation rate of $\kappa\sim0.05\omega_m$, a mechanical damping $\gamma_m\sim 10^{-4}\omega_m$, and coupling strengths $\chi_1\sim0.01\omega_m$ and $\chi_2\sim0.03\omega_m$, we have $\bar{n}_{\rm th, max}\sim70$. This indicates that two-mode mechanical squeezing is robust against thermal fluctuations and ground-state precooling of the mechanical modes may not be necessary.

\section{Mechanical oscillators in separate single-mode cavities}
\begin{figure}[t]
\includegraphics[width=3.0in]{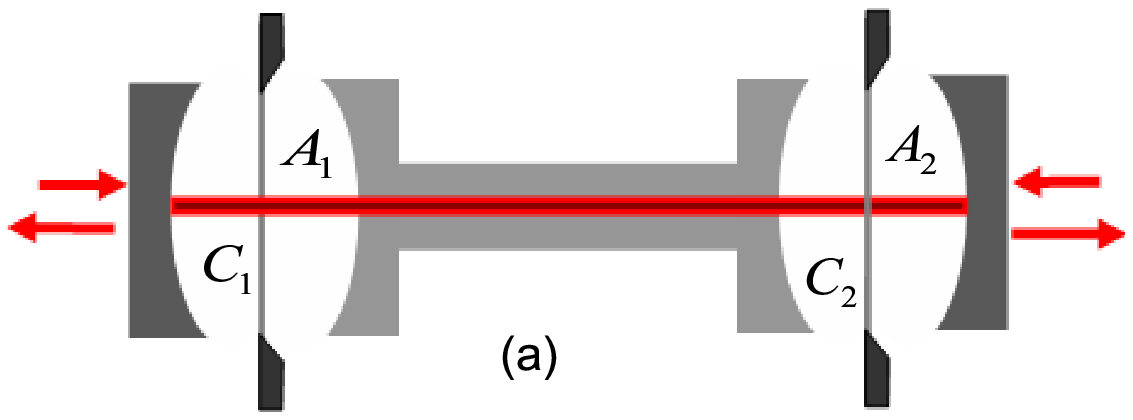}
\includegraphics[width=3.0in]{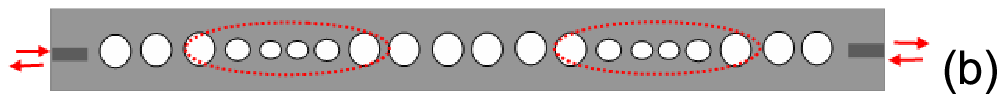}
\caption{(a) Two vibrating membranes ($C_j$)  in separate single-mode cavities ($A_j$)  are optically coupled by either an optical fiber. (b) Two evanescent-wave-coupled optomechanical crystal nanocavities.}
\end{figure}

In this section we show that the same dissipative approach can also be utilized to generate a two-mode squeezed state of two mechanical oscillators in separate single-mode cavities. The specific example that we consider consists of two identical single-mode cavities of frequency $\omega_c$, optically coupled to via fiber (a) or evanescent-wave coupling (b), see Fig.~4. Each cavity  field is driven by a modulated laser and optomechanically coupled to a mechanical oscillator of frequency $\omega_m$. Adopting the same symbols as before the Hamiltonian of that system reads
\begin{eqnarray}
H&=&\sum_{j~=1,2}\Big\{\omega_{c} \hat A_j^\dag \hat A_j+\omega_m \hat C_j^\dag \hat C_j+g \hat A_j^\dag \hat A_j(\hat C_j+\hat C_j^\dag)\nonumber\\
&+&i\mathcal{E}(t)e^{-i\omega_lt} \hat A_j^\dag-i\mathcal{E}^*(t)e^{i\omega_lt} \hat A_j\Big\}\nonumber\\
&+&\mathcal{J}_{12}(\hat A_1\hat A_2^\dag+\hat A_1^\dag \hat A_2), \label{hamil}
\end{eqnarray}
where $g$ account for the optomechanical coupling strengths, assumed to be identical, and $\mathcal{J}_{12}$ describes the coupling between the two cavities. Here we consider two pump fields of equal frequency $\omega_l$ and time-dependent amplitude
\begin{equation}
\mathcal{E}(t)={\cal E}_1e^{-i\Omega t+i\varphi_1}+{\cal E}_2e^{i\varphi_2}.
\end{equation}
For that symmetrical situation, both cavity fields have the same amplitude $\alpha_j(t)=\alpha(t)$ and the linearized Hamiltonian is
\begin{eqnarray}
H_{\rm lin}&=&\sum_j\Delta \hat a_j^\dag \hat a_j+\omega_m \hat c_j^\dag \hat c_j
+[\chi^*(t)\hat a_j+\chi(t)\hat a_j^\dag](\hat c_j+\hat c_j^\dag)\nonumber\\
&+&\mathcal{J}_{12}(\hat a_1\hat a_2^\dag+\hat a_1^\dag \hat a_2),
\end{eqnarray}
where
\begin{equation}
\chi(t)=g\alpha(t)
\end{equation}
and the detuning $\Delta\simeq (\omega_c-\omega_l)$ for the weak optomechanical coupling. By introducing the new bosonic operators
\begin{subequations}
\begin{align}
\hat b_{1,2}&=(\hat c_1\pm \hat c_2)/\sqrt{2},\\
\hat d_{1,2}&=(\hat a_1 \pm \hat a_2)/\sqrt{2},
\end{align}
\end{subequations}
the Hamiltonian separates as before into the sum of two uncoupled Hamiltonians, $H_{\rm lin}=\sum_{j=1,2}\tilde{H}_{j}^{\rm lin}$, where
\begin{eqnarray}
\tilde{H}_{j}^{\rm lin}&=\Delta_j \hat d_j^\dag \hat d_j+\omega_m\hat b_j^\dag \hat b_j
+[\chi^*(t)\hat d_j+\chi(t)\hat d_j^\dag](\hat b_j+\hat b_j^\dag),\nonumber\\
\end{eqnarray}
with effective detunings $\Delta_{1}=\Delta+\mathcal{J}_{12}$ and $\Delta_{2}=\Delta-\mathcal{J}_{12}$. The dynamics of the two independent optomechanical sub-systems are governed by the same master equations as before, see Eq.~(\ref{master1}).

In contrast to the preceding case, though, in the two-cavity setup it is not possible to simultaneously achieve the single-mode squeezing of the normal modes $b_1$ and $b_2$ since there is only one driving laser field and one set of control parameters ($\omega_l, \Omega, \varphi_j$). However, for sufficiently weak mechanical decoherence it is possible in principle to implement a two-step process that can still achieve that goal.

For the first step we choose the frequencies $\omega_l$ and $\Omega$  such that
\begin{eqnarray}
\label{cd2}
\Delta_1&=&\omega_c-\omega_l + {\cal J}_{12}=\omega_m,\nonumber \\
\Omega&=&2\omega_m,
\end{eqnarray}
and the phases
\begin{eqnarray}
\varphi_1&=&\phi_1-\arctan(\omega_m/\kappa)\nonumber \\
\varphi_2&=&\arctan(\omega_m/\kappa),
\end{eqnarray}
where $\phi_1$ is arbitrary. With the transformation $\hat a_j\rightarrow \hat a_je^{-i\Delta_jt}$ and $\hat b_j\rightarrow \hat b_je^{-i\omega_mt}$, the Hamiltonians $\tilde{H}_j^{\rm lin}$ reduce to
\begin{subequations}
\begin{align}
\label{hamil12}
\tilde{H}_1^{\rm lin}&=(\chi_1 e^{-i\phi_1} \hat b_1+\chi_2 \hat b_1^\dag)\hat d_1\nonumber\\
&~+(\chi_{2}e^{-2i\omega_m t} \hat b_1+\chi_1 e^{i(2\omega_m t-\phi_1)} \hat b_1^\dag)\hat d_1+\text {H.c.}\\
\tilde{H}_2^{\rm lin}&=(\chi_1 e^{i(2\mathcal{J}_{12}t-\phi_1)} \hat b_1+\chi_2e^{-2i\mathcal{J}_{12}t}\hat  b_1^\dag)\hat d_1\nonumber\\
&~+(\chi_1 e^{2i(\mathcal{J}_{12}-\omega_m)t} \hat b_1+\chi_{2}e^{i\left [2(\mathcal{J}_{12}+\omega_m )t-i\phi_1\right ]} \hat b_1^\dag)\hat d_1\nonumber\\
&~+\text {H.c.}
\end{align}
\end{subequations}
For $\chi_j\ll\{\omega_m,\mathcal{J}_{12}, |\mathcal{J}_{12}-\omega_m|\}$, the non-resonant terms in these Hamiltonians can be neglected and they reduce to $\tilde{H}_1^{\rm lin}\simeq(\chi_1 e^{-i\phi_1} \hat b_1+\chi_2 \hat b_1^\dag)\hat d_1+\text{H.c.}$ and $\tilde{H}_2^{\rm lin}\simeq0$, respectively. For the mode $b_1$ this is formally the same situation as encountered in Section IIB. Neglecting as before the mechanical damping, $\gamma_m=0$,  cavity dissipation brings likewise that normal mode into a steady-state single-mode squeezed vacuum for long enough time, and at the same time mode $b_2$ simply decays into the vacuum, i.e.,
\begin{subequations}
\begin{align}
&\tilde{\rho}_{\hat b_1}(t\ge t_{\rm min})=\hat S(\xi_1)|0_{b_1}\rangle\langle0_{b_1}|\hat S^\dag(\xi_1),\label{sqb12}\\
&\tilde{\rho}_{\hat b_2}(t\ge t_{\rm min})=|0_{b_2}\rangle\langle0_{b_2}|.
\end{align}
\end{subequations}
where $t_{\rm min}$, as defined before, is a time long enough that steady state has been reached. After the first step the mechanical oscillators $c_1$ and $c_2$ are therefore prepared in a pure two-mode squeezed state, although not a standard two-mode squeezed {\em vacuum}. That latter goal can be achieved in a second step by changing the frequency and phase of the pump laser at time $t_{\rm min}$ so that
\begin{eqnarray}
\label{cd3}
\Delta_2&=&\omega_c-\omega_l - {\cal J}_{12}=\omega_m,\nonumber \\
\Omega&=&2\omega_m,
\end{eqnarray}
and the phases
\begin{eqnarray}
\varphi_1&=& \phi_1-\arctan(\omega_m/\kappa) +\pi,\nonumber \\
\varphi_2&=&\arctan(\omega_m/\kappa).
\end{eqnarray}
For the weak optomechanical coupling, we then have that $\tilde{H}_1^{\rm lin}\simeq0$, which means that the mode $b_1$ evolves freely in the state of Eq.~(\ref{sqb12}), while $\tilde{H}_2^{\rm lin}\simeq(\chi_1 e^{-i\phi_2} \hat b_2+\chi_2 \hat b_2^\dag)\hat d_2+\text{H.c}$. After a time $t\ge2t_{min}$ both the normal modes $b_1$ and $b_2$ are therefore in steady-state single-mode squeezed vacua,
\begin{subequations}
\begin{gather}
\tilde{\rho}_{b_1}(t\ge2t_{\rm min})=\hat S(\xi_1)|0_{b_1}\rangle\langle0_{b_1}|\hat S^\dag(\xi_1),\\
\tilde{\rho}_{b_2}(t\ge2t_{\rm min})=\hat S(\xi_2)|0_{b_2}\rangle\langle0_{b_2}|\hat S^\dag(\xi_2),
\end{gather}
\end{subequations}
and the mechanical modes $c_1$ and $c_2$ are in the two-mode squeezed vacuum state,
\begin{equation}
\rho_{c_1c_2}(t\ge 2t_{\rm min})=\hat S_{12}(\xi_{12})|0_{c_1},0_{c_2}\rangle\langle0_{c_1},0_{c_2}| \hat S_{12}^\dag(\xi_{12}).
\end{equation}

When accounting for mechanical damping the two-step preparation scheme remains efficient for mean thermal phonon number such that $\gamma_m\bar{n}_{\text{th}}\ll\kappa$ so that $[\gamma_m\bar{n}_{\text{th}}]^{-1} \gg t_{\rm min}\gg \kappa^{-1}$ and thermal effects can be neglected during the state preparation. In case this condition is not satisfied, $\gamma_m\bar{n}_{\text{th}}\ge\kappa$, following the second step the mode $b_1$ is thermalized  while the  mode $b_2$ is in a squeezed thermal state,
\begin{subequations}
\begin{align}
&\tilde{\rho}_{b_1}(t\ge 2t_{min})=\tilde{\rho}_{{\rm th}, b_1},\label{thm}\\
&\tilde{\rho}_{b_2}(t\ge 2t_{min})=\hat S(\xi_2)\tilde{\rho}_{\text {th},b_2}\hat S^\dag(\xi_2),
\end{align}
\end{subequations}
with
\begin{eqnarray}
\varrho_{\text {th}, b_i}&=&\sum_{n_{b_i}=0}^{\infty}\frac{\bar{n}_i^{n_{b_i}}}{(\bar{n}_i+1)^{n_{b_i}+1}} |n_{b_i}\rangle\langle n_{b_i}|,\,\,\,\,\,i=1, 2\nonumber \\
 \xi_2&=&(r-\tilde{r}) e^{-i\phi},\nonumber \\
 \tilde{r}&=&\frac{1}{4}\ln\frac{2d_0(d_1+d_2)+1}{2d_0(d_1-d_2)+1},
 \end{eqnarray}
with the mean thermal excitation number $\bar{n}_1=\bar{n}_{\text{th}}$ and $\bar{n}_2=\sqrt{(d_0d_1+1/2)^2-d_0^2d_2^2}-1/2$. Note however that a two-step procedure is not required in that situation since a single step with laser parameters satisfying either Eq.~(\ref{cd2}) [(\ref{cd3})] will result in the preparation of mode $b_1$ [or $b_2$] in a squeezed thermal state while the other mode in the thermal state~(\ref{thm}), with the degree of EPR correlations between the mechanical modes $c_1$ and $c_2$
\begin{align}
\Delta_{\rm EPR, min}(\infty)=e^{-2r}(1-d_0)+(2\bar{n}_{\text{th}}+1)(1+d_0).
\end{align}
Hence, the maximum mean number  of thermal phonons $\bar{n}_{\rm th, max}$ for which squeezing can be achieved is
\begin{align}
\bar{n}_{\rm th, max}=\frac{1-d_0}{2(1+d_0)}(1-e^{-2r}),
\end{align}
which is obviously smaller than that in Eq.(\ref{ineq1}) because just one normal mode is now in a squeezed state and it is therefore harder to maintain quantum correlations.

\section{Conclusion}
In conclusion, we have proposed two possible quantum optomechanical setups to generate two-mode mechanical squeezed states of mechanical oscillators in optical cavities driven by modulated lasers. We showed analytically that for appropriate laser pump parameters the two oscillators can be prepared into a stationary two-mode mechanical squeezed vacuum with the aid of the cavity dissipation when choosing appropriate positions of the membranes in the optical cavities. The effect of thermal fluctuations on the two-mode mechanical squeezing was also investigated in detail, and we showed that mechanical squeezing is achievable without pre-cooling the mechanical oscillators to their quantum ground states. The present schemes are deterministic and can be implemented in a variety of optomechanical systems with current state-of-the-art experimental techniques.

\acknowledgements

This work is supported by the National Natural Science Foundation of China (Grant Nos.~11274134, 11074087 and 61275123),
the National Basic Research Program of China (Grant No. 2012CB921602),
the DARPA QuASAR and ORCHID programs through grants from AFOSR and ARO, the US Army Research Office, and the US National Science Foundation. THT is also supported by the CSC.

\end{document}